\documentclass[12pt,a4paper,titlepage]{article}
\usepackage{amsmath,color}
\usepackage{mathrsfs}

\def\slash#1{\not\!\!#1}
\renewcommand{\eqref}[1]{(\ref{#1})}

\usepackage{amsmath,amssymb}
\usepackage{graphicx}
\usepackage{here}

\definecolor{hyprf}{cmyk}{1,0.5,0,0}

\begin{document}

\thispagestyle{empty}

{\hbox to\hsize{
\vbox{\noindent March 2019 \hfill IPMU18-0198 }}
\noindent  revised version \hfill WU-HEP-18-12}

\noindent
\vskip1.5cm
\begin{center}

{\large\bf Polonyi-Starobinsky supergravity with inflaton in
\vglue.1in
               a massive vector multiplet with DBI and FI terms
}

\vglue.3in

Hiroyuki Abe~${}^{a}$, Yermek Aldabergenov~${}^{b,c}$, Shuntaro Aoki~${}^{a}$, \\
and  Sergei V. Ketov~${}^{d,e,f}$
\vglue.1in

${}^a$~Department of Physics, Waseda University, Tokyo 169-8555, Japan\\
${}^b$~Department of Physics, Faculty of Science, Chulalongkorn University,\\
Thanon Phayathai, Pathumwan, Bangkok 10330, Thailand \\
${}^c$~Institute of Experimental and Theoretical Physics, Al-Farabi Kazakh National 
University, 71 Al-Farabi Avenue, Almaty 050040, Kazakhstan\\
${}^d$~Department of Physics, Tokyo Metropolitan University, \\
1-1 Minami-ohsawa, Hachioji-shi, Tokyo 192-0397, Japan \\
${}^e$~Research School of High-Energy Physics, Tomsk Polytechnic University,\\
2a Lenin Avenue, Tomsk 634050, Russian Federation \\
${}^f$~Kavli Institute for the Physics and Mathematics of the Universe (IPMU),
\\The University of Tokyo, Chiba 277-8568, Japan \\
\vglue.1in

abe@waseda.jp, shun-soccer@akane.waseda.jp, Yermek.A@chula.ac.th, ketov@tmu.ac.jp
\end{center}

\vglue.3in

\begin{center}
{\Large\bf Abstract}
\end{center}
\vglue.1in
\noindent  We propose the Starobinsky-type inflationary model in the matter-coupled $N=1$ four-dimensional supergravity with the massive vector multiplet that has inflaton (scalaron) and goldstino amongst its field components, whose action includes the Dirac-Born-Infeld-type kinetic term and the generalized (new) Fayet-Iliopoulos-type term, without gauging the R-symmetry. The $N=1$ chiral matter ("hidden sector") is described by the modified Polonyi model needed for spontaneous supersymmetry breaking after inflation. We compute the bosonic action and the scalar potential of the model, and show that it can accommodate the positive (observed) cosmological constant (as the dark energy) and  the  spontaneous supersymmetry breaking at high scale after the Starobinsky inflation.

\newpage

\section{Introduction}

Supergravity is well motivated in theoretical high-energy physics of elementary particles beyond the Standard Model, in the gravitational theory, and in superstring theory. 
Supergravity theory is also considered as the promising framework  in theoretical cosmology, because it has the natural candidate for dark matter particle, known as the Lightest Supersymmetric Particle (LSP). It is also worth mentioning that the phenomenological supergravity model building, both for particle physics and cosmology, is highly non-trivial, because the supergravity interactions are very restrictive (due to local supersymmetry), and the low-energy bounds (given by the Standard Model of elementary particles and the $\Lambda$CDM Model of Cosmology) are tight.

The standard approach to the supergravity-based cosmology and the related inflationary model building employs {\it chiral} superfields for matter, inflaton and goldstino. The inflaton field is the real scalar field driving inflation, whereas a chiral multiplet has a complex physical scalar. Hence, another (non-inflaton) real scalar has to be stabilized during a single-field inflation. Since Supersymmetry (SUSY) is spontaneously broken during inflation, there must be also the goldstino (fermionic) field that is usually assigned to a chiral multiplet too. Then the input is provided by a (non-holomorphic) K\"ahler potential and a (holomorphic) superpotential of the chiral superfields involved. The inflationary scalar potential in supergravity generically suffers from the  so-called $\eta$-{\it problem}, so that the input should be carefully designed to avoid this problem --- see e.g., 
Refs.~\cite{Kawasaki:2000yn,Kallosh:2010ug,Kallosh:2010xz,Yamaguchi:2011kg,
Ketov:2014qha,Ketov:2014hya,Ketov:2015tpa} for many examples.  

The viable alternative to the standard approach in the inflationary cosmology based on supergravity is possible by employing a massive {\it vector} multiplet that has only one (real) physical scalar to be identified with inflaton, and its fermionic superpartner to be identified with goldstino (there is the massive vector field too). This alternative approach in its most minimal  form (without chiral matter) can accommodate cosmological inflation with any values of the Cosmic Microwave Background (CMB) radiation tilts $n_s$ and $r$ \cite{Farakos:2013cqa,Ferrara:2013rsa,Ferrara:2014cca}, but fails to achieve a positive cosmological constant (dark energy) and spontaneously broken SUSY {\it after} inflation. Phenomenological applications of supergravity to particle physics
(reheating) after inflation demand adding the "hidden sector" to be responsible for spontaneous SUSY breaking and, next, a mediation of the SUSY breaking from the hidden sector to the observable (low-energy) phenomena described by the Standard Model. The simplest candidate for the hidden sector is given by the so-called Polonyi model of a single chiral superfield  with a {\it linear} superpotential \cite{Polonyi:1977pj}. The Polonyi model was employed  for SUSY breaking in the supergravity model with inflaton in the massive vector multiplet in Refs.~\cite{Aldabergenov:2016dcu,Aldabergenov:2017bjt}. 

Starobinsky inflation \cite{Starobinsky:1980te} can be also realized in the supergravity with
inflaton (scalaron) in the massive vector multiplet, without chiral matter  
\cite{Farakos:2013cqa,Ferrara:2013rsa}. However, in the presence of the Polonyi
superfield, it was found to lead to instability of the inflationary trajectory 
\cite{Aldabergenov:2017hvp}. A cure to the last problem was also proposed in
Ref.~\cite{Aldabergenov:2017hvp} by modifying the embedding of the Starobinsky model into supergravity. The alternative possibility was proposed in Ref.~\cite{Abe:2018plc} by
 removing the Polonyi superfield, but adding the generalized Fayet-Iliopoulos (PI) term
instead, which does not require gauging the R-symmetry. Such new FI terms in supergravity
were recently introduced in Refs.~\cite{Cribiori:2017laj,Kuzenko:2018jlz,Antoniadis:2018oeh,Aldabergenov:2018nzd}. Then it is possible to get a positive cosmological constant via the D-type spontaneous SUSY breaking by the use of the FI term \cite{Antoniadis:2018oeh}. However, because of the tiny observed value of the cosmological constant, the
scale of such SUSY breaking appears to be very small and, hence, inappropriate for particle physics phenomenology.

In this paper we add the Polonyi superfield to the supergravity model of 
Ref.~\cite{Abe:2018plc}, in order to achieve a higher scale of spontaneous SUSY breaking
via the F-type SUSY breaking, while also keeping  the D-type SUSY breaking due to the FI term describing the observed positive cosmological constant.  Our construction appears to be very delicate and rather complicated, and is based on the following theoretical resources (tools):
\begin{itemize}
\item the manifest ({\it linearly} realized) local $N=1$ supersymmetry,
\item the inflaton (scalaron) and the goldstino in the massive $N=1$ vector multiplet,
\item the vector multiplet kinetic terms of the Dirac-Born-Infeld (DBI) structure  inspired by superstrings and D-branes \cite{Born:1934gh,Fradkin:1985qd,Leigh:1989jq},
\item the generalized Fayet-Iliopoulos (FI) term, without gauging the R-symmetry
\cite{Cribiori:2017laj,Kuzenko:2018jlz,Antoniadis:2018oeh,Aldabergenov:2018nzd},~\footnote{The standard FI term in supergravity is known to be very restrictive, that is problematic for cosmological applications \cite{Binetruy:2004hh}.}
\item the K\"ahler potential with the stabilizing (quartic) term, and the modified kinetic term
of the Polonyi superfield,
\item the Starobinsky (real) master function modified by a linear term in supergravity.
\end{itemize}
Our motivation and reasons for these modifications of the Polonyi-Starobinsky (PS) supergravity of 
 Refs.~\cite{Aldabergenov:2017hvp,Abe:2018plc} are explained in the main text of this paper.

Our paper is organized as follows. The technical setup based on the superconformal tensor calculus with a chiral compensator is briefly reviewed in Sec.~2. In Sec.~3 we define our supersymmetric action for the vector multiplet, and calculate its bosonic terms, including the auxiliary field and the scalar potential. In Sec.~4 we introduce the modified Polonyi model in the context of PS supergravity, and study its properties during and after inflation. In Sec.~5 we propose a more general action in the curved superspace of the "old-minimal" supergravity, connect it to the actions defined in Secs.~2 and 3, and use  the superfield formulation to verify our results found in the superconformal tensor calculus approach.~\footnote{Though the superspace approach and the superconformal tensor calculus approach to supergravity are equivalent, their correspondence is non-trivial.} 
 Our conclusion is given by Sec.~6. Technical details are collected in Appendices A, B, C and D. 

\section{Superconformal tensor calculus}

We use the conformal $N=1$ supergravity techniques~\cite{Kaku:1978nz,Kaku:1978ea,Townsend:1979ki, Kugo:1982cu,Kugo:1983mv}, and follow the notation and conventions of Ref.~\cite{Freedman:2012zz}. In addition to the symmetries of Poincar\'e supergravity, one also has the gauge invariance under dilatations, conformal boosts and
$S$-supersymmetry, as well as under $U(1)_A$ rotations. The gauge fields of dilatations and   $U(1)_A$ rotations are denoted by $b_{\mu}$ and $A_{\mu}$, respectively. A multiplet of conformal supergravity has charges with respect to dilatations and $U(1)_A$ rotations, called Weyl and chiral weights, respectively, which are denoted by the pair $({\rm{Weyl\ weight, chiral\ weight}})$ in what follows. 

A chiral multiplet has field components 
\begin{align}
S=\{S,P_L\chi, F\}, \label{chiral}
\end{align}
where $S$ and $F$ are complex scalars, and $P_L\chi$ is a left-handed Weyl fermion ($P_L$ is the chiral projection operator).  In this paper, we use two types of chiral multiplets: the conformal compensator $S_0$ and the matter multiplets $S^i$, where the index $i=1, 2 ,3 ,\ldots$, counts the matter multiplets.  The $S_0$ has the weights $(1,1)$ and is used to fix some of the superconformal symmetries. The matter multiplets $S^i$ have the weights $(0,0)$. The anti-chiral multiplets are denoted by $\bar{S}_0$ and $\bar{S}^{\bar{i}}$. 

As regards a general (complex or real) multiplet, it has  the field content
\begin{align}
\Phi=\{ \mathcal{C}, \mathcal{Z}, \mathcal{H}, \mathcal{K}, \mathcal{B}_{a}, \Lambda , \mathcal{D}\} , \label{general}
\end{align}
where $ \mathcal{Z}$ and $\Lambda $ are fermions, $\mathcal{B}_a$ is a (complex or real) vector, and others are (complex or real) scalars, respectively.

The (gauge) field strength multiplet $W$ has the weights $(3/2,3/2)$ and the following field components: 
\begin{align}
\bar{\eta } W=\left\{ \bar{\eta }P_L\lambda ,\frac{1}{\sqrt{2}}\left(  -\frac{1}{2}\gamma _{ab}\hat{F}^{ab}+iD\right) P_L\eta  ,\bar{\eta  } P_L\slash{D}\lambda \right\} ,
\end{align}
where $\eta $ is the dummy spinor, $\hat{F}_{ab}=\partial_aB_b-\partial_bB_a+\bar{\psi}_{[a}\gamma_{b]}\lambda \equiv F_{ab}+\bar{\psi}_{[a}\gamma_{b]}\lambda$ is the  superconformally covariant field strength, the $\psi_a$ is gravitino, the $\lambda $ and $D$ are Majorana fermion and the real auxiliary scalar, respectively. The related expressions of the multiplets $W^2$ and $W^2\bar{W}^2$, which are embedded into the chiral multiplet~$\eqref{chiral}$ and the general multiplet~$\eqref{general}$, respectively, are 
\begin{align}
&W^2=\left\{ \cdots , \cdots, \cdots+\frac{1}{2} (FF-F\tilde{F})-D^2\right\} ,\\
&W^2\bar{W}^2=\left\{ \cdots , \cdots, \cdots,\cdots , \cdots, \cdots, \cdots+\frac{1}{2}| (FF-F\tilde{F})-2D^2|^2\right\} ,
\end{align}
where we have omitted the fermionic terms (denoted by dots) for simplicity. 
In addition, we use the book-keeping notation  $FF=F_{ab}F^{ab}$ and $\tilde{F}^{ab}\equiv -\frac{i}{2}\varepsilon ^{abcd}F_{cd}$ throughout the paper.

We also need another chiral multiplet
\begin{align}
\nonumber &\Sigma \left( \bar{W}^2/|S_0|^4\right) = \\
& \left\{ -\frac{ (\frac{1}{2}FF+\frac{1}{2}F\tilde{F}-D^2)}{|S_0|^4} +\cdots , \cdots, \frac{F_0}{|S_0|^4S_0}(FF+F\tilde{F}-2D^2)+\cdots\right\} ,  \label{D2W2}
\end{align}
where $\Sigma$ is the chiral projection operator \cite{Kugo:1982cu, Kugo:1983mv}. The argument of $\Sigma$ requires specific Weyl and chiral weights: in order for $\Sigma \Phi$ to make sense, $\Phi$ must satisfy $w-n=2$, where $(w,n)$ are the Weyl and chiral weights of $\Phi$. We adjust the correct weights of the argument, by inserting the factor $|S_0|^4$. Equation~$\eqref{D2W2}$ is the conformal  supergravity counterpart of the superfield $\bar{D}^2\bar{W}^2$. 

The covariant derivative of $W$ is given by \cite{Kugo:1983mv}
\begin{align}
\mathcal{D}W=\{ -2D, \cdots , \cdots, \cdots,\cdots , \cdots,\cdots \} 
\end{align}
of the weights $(2,0)$. Here, the dots in the higher components also include some bosonic terms, but we do not write down them here for simplicity (see Ref.~\cite{Cribiori:2017laj} for their explicit expressions).

A massive vector multiplet $V$ has the field components 
\begin{align}
V=\{C,Z,H,K,B_a, \lambda ,D\}~,
\end{align}
while all of them are either real (bosonic) or Majorana (fermionic). The weights of $V$ are $(0,0)$.

The bosonic parts of the F-term invariant action formulas are 
\begin{align}
[S]_F=\int d ^{4}x\sqrt{-g}\frac{1}{2}\left( F+\bar{F}\right), \label{Fformula}
\end{align}
while they can be applied only when $S$ has the weights $(3,3)$. The bosonic part of the D-term formula for a real multiplet $\phi $ of the weights $(2,0)$ is 
\begin{align}
 [\phi ]_D= \int d^{4}x\sqrt{-g}\left( D_{\phi}-\frac{1}{3}C_{\phi}R(\omega)\right) ,\label{Dformula}
\end{align}
where $R(\omega)$ is the superconformal Ricci scalar in terms of spacetime metric and $b_{\mu}$~\cite{Freedman:2012zz}. The  $C_{\phi}$ and $D_{\phi}$ are the first and the last components of $\phi$, respectively.


\section{Vector multiplet coupled to chiral matter }

Let us consider a (chiral) matter coupled extension of the DBI and FI system investigated in Ref.~\cite{Abe:2018plc}, with the action
\begin{align}
S=S_{V,M}+S_{DBI}+S_{FI}, \label{total action}
\end{align}
where we have introduced
\begin{align}
&S_{V,M}=\biggl[ |S_0|^2\mathcal{H}\biggr] _D+2[S_0^3\mathcal{W}]_F, \label{S_VM} \\
&S_{DBI}=-\frac{1}{2}[W^2]_F +\biggl[  \frac{\alpha W^2\bar{W}^2}{|S_0|^4\left( 1-2\alpha \mathcal{A}+\sqrt{1-4\alpha \mathcal{A}+4\alpha ^2\mathcal{B}^2} \right)}\biggr] _D ,\label{S_DBI}\\
&S_{FI}=\biggl[ |S_0|^2\mathcal{H}\mathcal{I}\frac{W^2\bar{W}^2}{(\mathcal{D}W)^2(\bar{\mathcal{D}}\bar{W})^2}\mathcal{D}W\biggr] _D . \label{S_FI}
\end{align}
Here, $\mathcal{H}$, $\mathcal{I}$, and $\alpha $ are arbitrary weightless real functions of $V, S^i$, and $\bar{S}^{\bar{i}}$, the $\mathcal{W}$ is a holomorphic superpotential that depends on $S^i$ only. The $\mathcal{A}$ and $\mathcal{B}$ in $S_{DBI}$ are given by 
\begin{align}
\mathcal{A}=\bar{\Sigma }\left( \frac{W^2}{|S_0|^4} \right) +{\rm{h.c.}}, \ \ \mathcal{B}=\bar{\Sigma }\left( \frac{W^2}{|S_0|^4} \right) -{\rm{h.c.}}
\end{align} 
More general actions are defined in Sec.~5 and Appendix~\ref{detail_DBI}. 

The action $\eqref{total action}$ reduces to the action of Ref.~\cite{Aldabergenov:2016dcu} in the limit of the vanishing $\alpha$ and $\mathcal{I}$. It is remarkable that one can introduce a  superpotential without any restriction because the new FI term does not require the gauged 
R-symmetry. Under K\"ahler transformations, the $S_0$, $\mathcal{H}$, and $\mathcal{W}$ transform as 
\begin{align}
S_0\rightarrow S_0e^{Y}, \ \ \mathcal{H} \rightarrow \mathcal{H}e^{-Y-\bar{Y}}, \ \ \mathcal{W}\rightarrow \mathcal{W}e^{-3Y},\label{KT}
\end{align} 
where $Y$ is a holomorphic function of $S^i$. That is why we inserted the factor 
$\mathcal{H}$ in Eq.~$\eqref{S_FI}$ for the K\"ahler invariance, as was pointed out in Refs.~\cite{Antoniadis:2018oeh,Aldabergenov:2018nzd}.

To obtain the Lagrangian in field components, we have to eliminate the unwanted symmetries of the conformal supergravity by gauge fixing, and solve for the auxiliary fields by their (algebraic) equations of motion. These steps can be done separately in Eqs.~$\eqref{S_VM}$-$\eqref{S_FI}$, except for the integration of the auxiliary field $D$. Let us focus on the $S_{V,M}$ first, and demonstrate the result for the bosonic part only. The details of derivation are given in Appendix~\ref{detail_VM}. The resulting Lagrangian $\mathcal{L}_{V,M}$ reads~\footnote{The Lagrangian density is defined by $S=\int d^4 x \sqrt{-g}\mathcal{L}$.}
\begin{align}
  \mathcal{L}_{V,M}=&\frac{1}{2}R-\frac{1}{2}\mathcal{J}_{CC}(\partial_aC)^2 -\frac{1}{2}\mathcal{J}_{CC}B_a^2+\mathcal{J}_CD\\
 &-2\mathcal{J}_{i\bar{j}}\partial_aS^i\partial^a\bar{S}^{\bar{j}}+\left(-\mathcal{J}_{Ci}\partial_aC\partial^aS^i -i\mathcal{J}_{Ci}B_a\partial^aS^i+{\rm{h.c.}}\right)-V_F~,\label{L_VM}
\nonumber \end{align}
where we have defined
\begin{align}
\mathcal{J}=-\frac{3}{2}\log \left( -\frac{2}{3}\mathcal{H}\right),
\end{align}
and the subscripts on $\mathcal{J}$ and $\mathcal{W}$ denote the derivatives with respect to $C$ and $S^i(\bar{S}^{\bar{i}})$. The $V_F $ is the F-type scalar potential, 
\begin{align}
\nonumber V_F=&e^{2\mathcal{J}}\biggl[2\Delta^{i\bar{j}}\biggl\{\frac{1}{2}\mathcal{W}_{i}+\left(\mathcal{J}_{i}-\frac{\mathcal{J}_{Ci}\mathcal{J}_C}{\mathcal{J}_{CC}}\right) \mathcal{W}\biggr\}\biggl\{\frac{1}{2}\bar{\mathcal{W}}_{\bar{j}}+\left(\mathcal{J}_{\bar{j}}-\frac{\mathcal{J}_{C\bar{j}}\mathcal{J}_C}{\mathcal{J}_{CC}}\right) \bar{\mathcal{W}}\biggr\}\\
&+2\frac{\mathcal{J}_C^2}{\mathcal{J}_{CC}}|\mathcal{W}|^2-3|\mathcal{W}|^2\biggr] ,
\end{align}
where we have used the notation
\begin{align}
\Delta^{i\bar{j}}=\left(\Delta_{i\bar{j}}\right)^{-1}, \ \ \Delta_{i\bar{j}}=\mathcal{J}_{i\bar{j}}-\frac{\mathcal{J}_{Ci}\mathcal{J}_{C\bar{j}}}{\mathcal{J}_{CC}}~~.
\end{align}

Under the superconformal gauge fixing conditions~$\eqref{D_gauge}$-$\eqref{K_gauge}$,  the $S_{DBI}$ becomes (see also Appendix~\ref{detail_DBI}) 
\begin{align}
\mathcal{L}_{DBI}=&\frac{e^{4\mathcal{J}/3}}{8\alpha}\biggl[ 1- \sqrt{1-8\alpha e^{-4\mathcal{J}/3}\left( D^2-\frac{1}{2}FF\right)+4\alpha^2e^{-8\mathcal{J}/3}(F\tilde{F})^2}\biggr]. \label{LDBI}
\end{align}

The bosonic part of $\mathcal{L}_{FI}$ takes the same form as that in Ref.~\cite{Abe:2018plc}, with
\begin{align}
\mathcal{L}_{FI}=\mathcal{I}\frac{\left(  D^2-\frac{1}{2}FF\right)^2-\frac{1}{4}(F\tilde{F})^2}{6D^3}~, \label{LFI}
\end{align}
though with the matter-dependent $\mathcal{I}$-function in general.

Having obtained the total Lagrangian $\mathcal{L}=\mathcal{L}_{V,M}+\mathcal{L}_{DBI}+\mathcal{L}_{FI}$ before integrating out $D$, we can discuss the elimination of $D$ . The  $D$ equation of motion gives
\begin{align}
&\mathcal{J}_C +\frac{1}{\sqrt{\mathcal{F}^2-8\alpha e^{-4\mathcal{J}/3}D^2}}D+\frac{\mathcal{I}}{6}\left(  1+\frac{FF}{D^2}-\frac{3}{4}\frac{(FF)^2-(F\tilde{F})^2}{D^4}\right) =0, \label{EOM D}
\end{align}
where we have used the notation
\begin{align}
\mathcal{F}(F)\equiv \sqrt{1+4\alpha e^{-4\mathcal{J}/3}FF+4\alpha ^2e^{-8\mathcal{J}/3}(F\tilde{F})^2}~~.
\end{align}

Similarly to Ref.~\cite{Abe:2018plc}, we find a perturbative solution with respect to the FI term, 
\begin{align}
D=D^{(0)}+\mathcal{I}D^{(1)}+ \mathcal{O}(\mathcal{I}^2)~, \label{sol per}
\end{align}  
where we have introduced
\begin{align}
D^{(0)}&= K\mathcal{F}, \label{sol 0} \\
D^{(1)}&=-\frac{K^3}{6\mathcal{J}_C^3}\mathcal{F}\left( 1+\frac{FF}{K^2\mathcal{F}^2} -\frac{3}{4}\frac{(FF)^2-(F\tilde{F})^2}{K^4\mathcal{F}^4} \right)~, \label{sol 1}
\end{align}
and
\begin{align}
K^2\equiv \frac{\mathcal{J}_C^2}{1+8\alpha \mathcal{J}_C^2 e^{-4\mathcal{J}/3}}~~.
\end{align}

Substituting the solution~$\eqref{sol per}$ back into the Lagrangian, we obtain the  Lagrangian in the first order with respect to  $\mathcal{I}$ as follows:
\begin{align}
 \mathcal{L}=&\frac{1}{2}R-\frac{1}{2}\mathcal{J}_{CC}(\partial_aC)^2 -\frac{1}{2}\mathcal{J}_{CC}B_a^2-2\mathcal{J}_{i\bar{j}}\partial_aS^i\partial^a\bar{S}^{\bar{j}}\\
\nonumber &+\left(-\mathcal{J}_{Ci}\partial_aC\partial^aS^i-i\mathcal{J}_{Ci}B_a\partial^aS^i+{\rm{h.c.}}\right)-V_F\\
 &+\frac{e^{4\mathcal{J}/3}}{8\alpha } \left( 1\pm \frac{\mathcal{J}_C}{K} \mathcal{F}\right) \pm  \frac{\mathcal{I}}{6}K\mathcal{F} \left( 1-\frac{FF}{K^2\mathcal{F}^2} +\frac{(FF)^2-(F\tilde{F})^2}{4K^4\mathcal{F}^4}\right)+\mathcal{O}(\mathcal{I}^2).
 \nonumber \label{onshell_action}
\end{align}

When $B_{a}=0$, Eq.~$\eqref{EOM D}$ can be solved exactly, and its solution is given by
\begin{align}
D= \sqrt{\frac{\left( \mathcal{J}_C+\frac{\mathcal{I}}{6} \right)^2}{1+8\alpha e^{-4\mathcal{J}/3}\left( \mathcal{J}_C+\frac{\mathcal{I}}{6} \right)^2 }}~~.
\end{align}

Thus, the Lagrangian becomes 
\begin{align}
\nonumber \mathcal{L}=&\frac{1}{2}R-\frac{1}{2}\mathcal{J}_{CC}(\partial_aC)^2 -2\mathcal{J}_{i\bar{j}}\partial_aS^i\partial^a\bar{S}^{\bar{j}}\\
&+\left(-\mathcal{J}_{Ci}\partial_aC\partial^aS^i+{\rm{h.c.}}\right)-V_F-V_D ~~,\label{action_scalar}
\end{align} 
where we have
\begin{align}
V_D= -\frac{e^{4\mathcal{J}/3}}{8\alpha } \left( 1- \sqrt{1+8\alpha e^{-4\mathcal{J}/3}\left( \mathcal{J}_C+\frac{\mathcal{I}}{6} \right)^2} \right) .
\end{align}

The field dependent kinetic term of the vector field $B_a$ can be read off from Eq.~$\eqref{onshell_action}$, and it should be negative to avoid a ghost mode, i.e.
\begin{align}
-\frac{1}{4K} \left( \mathcal{J}_C -\frac{2}{3}\mathcal{I} +\frac{4}{3}\alpha K^2\mathcal{I}e^{-4\mathcal{J}/3}\right) <0. \label{cond_FF}
\end{align} 


\section{Inflation and SUSY breaking}\label{ISB}

Let us apply the supergravity model constructed in the previous section to a description of cosmological inflation, spontaneous SUSY breaking after inflation, and dark energy (positive cosmological constant). In this section we restore the (reduced) Planck mass $M_{\rm{P}}$ and the gauge coupling constant $g$.~\footnote{The vector multiplet gets its mass via Higgs effect, see Sec.~5 for more.}

We specify our model for the Starobinsky-type inflation by identifying the real scalar $C$ of the massive vector multiplet with the inflaton (Starobinsky scalaron). We also add the
 SUSY breaking (hidden) sector $\Phi$ as the modified Polonyi model, with
\begin{align}
\mathcal{J}=&J(C)+\frac{1}{2}|\Phi|^2+\zeta|\Phi|^4+\epsilon C(\Phi+\bar{\Phi}) ,\label{Kahler}\\
\mathcal{W}=&\mu (\Phi +\beta) ,\label{SP}
\end{align}
where the function $J$ is taken in the Starobinsky form  \cite{Ferrara:2013rsa,Ferrara:2014cca} modified by 
the linear term with the real coefficient $\gamma$ as follows:
\begin{align}
J=-\frac{3}{2}M_{\rm{P}}^2\log \left(  -\frac{C}{M_{\rm{P}}}e^{C/M_{\rm{P}}}\right)+\gamma M_{\rm{P}}C.\label{Starobinsky}
\end{align} 

Some comments are in order. The original Polonyi model \cite{Polonyi:1977pj}
is obtained in the case of the vanishing parameters $\zeta$, $\epsilon$ and $\gamma$, with the $\mu$ and $\beta$ as the real parameters. We have added the quartic coupling 
with the $\zeta$ parameter to the K\"ahler potential, and the mixing of the kinetic terms of $C$ and $\Phi$ with the parameter $\epsilon$ in \eqref{Kahler}. As regards 
Eq.~$\eqref{Starobinsky}$, we have also added the linear term in $C$. As is demonstrated below, all these modifications are necessary to achieve our goals given in the Abstract and Sec.~1.

The $F$- and $D$- type scalar potentials are given by 
\begin{align}
\nonumber V_F=&e^{2\mathcal{J}/M^2_{\rm{P}}}\mu^2\biggr[ \frac{1}{\Delta} |1+\frac{1}{M^2_{\rm{P}}}(\Phi+\beta)(\bar{\Phi}+4\zeta\bar{\Phi}|\Phi|^2+2\epsilon C-2\epsilon Q)|^2\\
&-\frac{3}{M^2_{\rm{P}}}|\Phi+\beta|^2P\biggr], \label{VF1} \\
V_D=&\frac{M_{\rm{P}}^4}{a^4}\left(  \sqrt{1+a^4 \left( \frac{g\mathcal{J}_C}{M_{\rm{P}}}+\frac{b}{6}  \right)^2} -1\right),  \label{VD1}
\end{align} 
where we have defined
\begin{align}
&\Delta =1+8\zeta |\Phi|^2-2\frac{\epsilon}{J_{CC}}~~,\\
&P=1-\frac{2}{3M^2_{\rm{P}}}\frac{\mathcal{J}_C^2}{J_{CC}}~~,\\
&Q=\frac{\mathcal{J}_C}{J_{CC}}~~,
\end{align}
with the dimensionless parameters $a$ and $b$ according to
\begin{align}
\alpha =\frac{a^4e^{4\mathcal{J}/3M_{\rm{P}}^2}}{8M_{\rm{P}}^4} \quad {\rm and} \quad  \mathcal{I}= b M_{\rm{P}}^2. \label{simplify}
\end{align} 

The kinetic term of the inflaton $C$ is given by
\begin{align}
-\frac{3M^2_{\rm{P}}}{4C^2}(\partial_aC)^2~~.
\end{align} 
It should be noticed that the $\gamma$-modification above does not affect this kinetic term. Hence, we define the "canonical" inflaton (scalaron) as~\footnote{Strictly speaking, due to the kinetic mixing between $C$ and $\Phi$, the scalar $\varphi$ is not canonical. However,
we can ignore this by keeping the parameter $\epsilon$ small, $|\epsilon|\ll 1$.}
\begin{align}
C/M_{\rm{P}}=-e^{\sqrt{\frac{2}{3}}\frac{\varphi }{M_{\rm{P}}} }~~.
\end{align} 


\subsection{During inflation}

In this subsection, we discuss the stabilization of $\Phi$ during inflation. Expanding the scalar potential around $\Phi=0$ up to the second order, we obtain
\begin{align}
V_F=V_0+\delta_R \Phi_R+\frac{1}{2}m^2_R\Phi_R^2+\frac{1}{2}m^2_I\Phi_I^2+\ldots, \label{Vinf}
\end{align} 
where $\Phi\equiv \frac{1}{\sqrt{2}}(\Phi_R+i\Phi_I)$ and
\begin{align}
&V_0\sim \frac{\mu^2 \beta^2}{12M_{\rm{P}}^2}e^{(3-2\gamma)e^x-x}(3-2\gamma)^2(4-\epsilon)+\frac{M_{\rm{P}}^4}{a^4}(R-1),\label{V0}\\
&\delta_R \sim -\frac{\mu^2 \beta^2}{3\sqrt{2}M_{\rm{P}}^3}e^{(3-2\gamma)e^x}(3-2\gamma)^2\epsilon (4-\epsilon)-\frac{3}{\sqrt{2}R}g^2M_{\rm{P}}^3\epsilon \left(1-\frac{2\gamma}{3}-\frac{b}{9g}\right) ,\label{delta_R}\\
&m^2_R\sim  -\frac{\mu^2 \beta^2\zeta }{2M_{\rm{P}}^2}e^{(3-2\gamma)e^x-3x}(3-2\gamma)^2+\frac{2g^2M^2\epsilon ^2}{R^3},\label{m_R}\\
&m^2_I\sim -\frac{\mu^2 \beta^2\zeta }{2M_{\rm{P}}^2}e^{(3-2\gamma)e^x-3x}(3-2\gamma)^2,\\
&x\equiv \sqrt{\frac{2}{3}}\frac{\varphi}{M_{\rm{P}}},\\
&R\equiv \sqrt{1+\frac{9}{4}a^4g^2\left(1-\frac{2\gamma}{3}-\frac{b}{9g}\right)^2}~~.
\end{align} 
Here we have assumed $x \gg 1$ and $M_{\rm{P}}^2|\zeta| \gg 1$,\footnote{To avoid a tachyonic mass, the $\zeta$ should be negative.} and have extracted only the dominant terms. The terms proportional to $\mu$ ($g$) come from $V_F$ ($V_D$), so that we call 
them as the $F$-($D$-) contributions, respectively.   

Let us comment on the role of $\gamma$ in Eq.~$\eqref{Starobinsky}$. As can be seen from Eq.~$\eqref{V0}$, the first term spoils the flatness of the scalar potential without $\gamma $. This fact was already pointed out in Ref.~\cite{Aldabergenov:2017hvp}. Hence, we have to assume $\gamma \geq 3/2$ from now on.~\footnote{Should $\mu^2$ be much smaller than the inflationary scale $\sim V_D$, we could neglect the first term in $V_0$ even when $\gamma < 3/2$. However, the value of $\mu$ comparable with the observed dark energy scale (the cosmological constant) implies the very small SUSY breaking scale in the vacuum, so that we do not consider this situation here.} 

For $\gamma \geq 3/2$, the $D$-term contributions are dominant in Eqs.~$\eqref{V0}$ and~$\eqref{delta_R}$. For the mass term in Eq.~$\eqref{m_R}$, however, the dominant term depends on the value of $\zeta$. Hence, we have to assume a relatively large $|\zeta|$ that strongly stabilizes $\chi_R$ at its origin. This requires $|\zeta| \gg \frac{g^2M_{\rm{P}}^2}{\mu^2}e^{-(3-2\gamma)e^x+3x}$. Thus, we obtain $\langle \Phi_R \rangle \sim \langle \Phi_I \rangle \sim 0$ during inflation. In the case of Starobinsky inflation, the $x$ varies between $5.5$ and $0.5$ according to \cite{Aldabergenov:2018qhs}, which implies a huge $|\zeta|$ in general. However, when $\gamma=3/2+\delta$ is very close to $3/2$, e.g., with $\delta \leq 0.001$, it is enough to require $|\zeta|M_{\rm{P}}^2 \gg \frac{g^2M_{\rm{P}}^4}{\mu^2}\times 10^7$.

If $\zeta=0$, the $D$-term contribution is dominant in Eq.~$\eqref{m_R}$. Then, $\chi_R$ tends to be light $(\sim g^2M_{\rm{P}}^2)$, and the deviation of $\langle \Phi_R \rangle $ from its origin is typically given by $\mathcal{O}(M_{\rm{P}})$. In this case, the isocurvature perturbations should also be included into consideration.

After integrating out $\Phi$, we obtain the effective potential during inflation as  
\begin{align}
V_{eff} =V_D|_{\Phi=0}~. \label{V_eff}
\end{align} 
This potential is known to be viable for inflation --- see Ref.~\cite{Abe:2018plc} for details.

\begin{figure}[H]
\includegraphics[width=12.0cm]{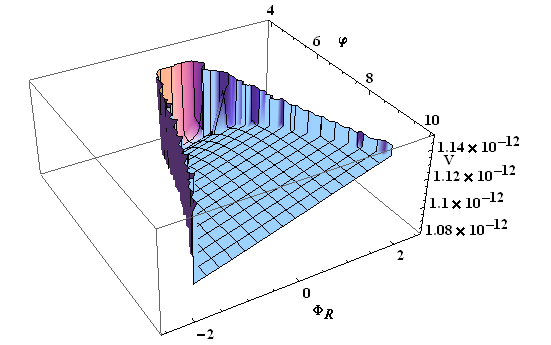}
\caption{3D plot of the potential $V$ in the $(\varphi, \Phi_R)$ plane with the parameters 
$\beta/M_{\rm{P}}=-1, \delta=0.1,\epsilon =-5\times 10^{-3},g=10^{-6},a=1,\mu=10^{-3},b=-6\gamma g,$ and $\zeta=10^2$. }
\label{fig_potential}
\end{figure}

One may wonder whether the expansion of the scalar potential of Polonyi superfield up to the second order in Eq.~$\eqref{Vinf}$ can be trusted. The full scalar potential $V$ given by a sum of  
Eqs.~\eqref{VF1} and \eqref{VD1} is dictated by Eqs.~$\eqref{Kahler}$ and $\eqref{SP}$. We  checked the stabilization numerically (Figure~\ref{fig_potential}) with the ratio $\beta/M_{\rm P}=-1$ for the relevant values of $\varphi$.

\subsection{After inflation}

In this subsection, we investigate the vacuum structure after inflation in our model. We expand the scalar potential around $\varphi=0$ and $\Phi=0$,~\footnote{Due to $\zeta\neq 0$, the value $ \langle \Phi \rangle =(\sqrt{3}-1)M_{\rm{P}}$ of the original Polonyi model does not correspond to a minimum~\cite{Buchmuller:2014pla}.}
\begin{align}
V=V_0+\delta_{\varphi} \varphi+\delta_R \Phi_R+\frac{1}{2}m^2_{\varphi} \varphi^2+m^2_{R\varphi}\Phi_R \varphi+\frac{1}{2}m^2_R\Phi_R^2+\frac{1}{2}m^2_I\Phi_I^2+\ldots, \label{EP}
\end{align} 
where  we have 
\begin{align}
&V_0 \sim \mathcal{O}(\mu^2),\\
&\delta_{\varphi}\sim \mathcal{O}(\mu^2/M_{\rm{P}}) ,\\
&\delta_R \sim \mathcal{O}(\mu^2/M_{\rm{P}}),\\
&m^2_{\varphi}\sim \mathcal{O}(\mu^2/M_{\rm{P}}^2),\\
&m^2_R\sim m^2_I\sim \mathcal{O}(\mu^2\zeta).
\end{align}
The explicit values are shown in Appendix~\ref{explicit_mass}. Also, we neglected the $D$-term contributions. This is valid when the SUSY breaking scale $\mu$ is much larger than the inflation scale $M_{\rm{P}}^2g$.\footnote{The $g$ is set to be $\sim 10^{-6}$ by the amplitude of the CMB power spectrum~\cite{Abe:2018plc}.}

For large $|\zeta|$, we obtain the following vacuum expectation values:
\begin{align}
 \langle \varphi \rangle \sim -\frac{\delta_{\varphi}}{m^2_{\varphi}}\sim \mathcal{O}(M_{\rm{P}}), \ \ \langle \Phi_R \rangle \sim \mathcal{O}(1/M_{\rm{P}}\zeta),\ \  \langle \Phi_I \rangle \sim 0. \label{vev_vacuum}
\end{align}
Inserting them into Eq.~$\eqref{EP}$ and neglecting the terms suppressed by $\zeta$, we find the minimum of the potential at
\begin{align}
\nonumber \langle V \rangle &\sim \langle V_F \rangle\\
\nonumber &\sim V_0-\frac{\delta_{\varphi}^2}{2m^2_{\varphi}}\\
 &\sim \frac{\mu^2 e^{3-2\gamma}}{3 M^2 (4 \epsilon -3)}\mathcal{V}(\beta,\gamma,\epsilon). \label{vev_VF}
\end{align}
Here $\mathcal{V}$ is a complicated function of $(\beta,\gamma,\epsilon)$ that can be read off from Eqs.~$\eqref{CC}$ and $\eqref{mass_inflaton}$. Like the original Polonyi model, we demand $\langle V_F \rangle =0$ by tuning the parameter $\beta$. 

Note that in the special case $\epsilon=\gamma=0$, we have
\begin{align}
\mathcal{V}(\beta,0,0)=3 \left(1-\frac{3 \beta ^2}{M_{\rm{P}}^2}\right),
\end{align}
so that $\beta =M_{\rm{P}}/\sqrt{3}$ leads to $\langle V_F \rangle =0$~\cite{Buchmuller:2014pla}.

Now $\gamma$ satisfies $\gamma >3/2$ and, therefore, the value of $\beta$ to obtain $\mathcal{V} =0$ is changed.~\footnote{The special case with $\gamma=3/2$ and $\epsilon=\zeta=0$ yielding a Minkowski vacuum is considered in Appendix~\ref{Minkvac}.} Unfortunately, we found that there is no real solution $\beta$ of $\mathcal{V}=0$, when $\epsilon =0$. One can check it by explicitly solving the equation $\mathcal{V}(\beta,\gamma ,0)=0$ that has the following four solutions:
\begin{align}
&\beta/M_{\rm{P}}=\pm \sqrt{\frac{\Lambda \pm\sqrt{\Sigma}}{\Gamma}}~,\label{sol_zero}\\
&\Gamma =32 \gamma ^6-96 \gamma ^5+32 \gamma ^4+240 \gamma ^3-414 \gamma ^2+162 \gamma +81,\\
&\Lambda =48\gamma^3-42\gamma^2+54,\\
&\Sigma=128 \gamma ^6-576 \gamma ^5+1120 \gamma ^4-1296 \gamma ^3+900 \gamma ^2-324 \gamma +81.
\end{align}

\begin{figure}[H]
\includegraphics[width=10.0cm]{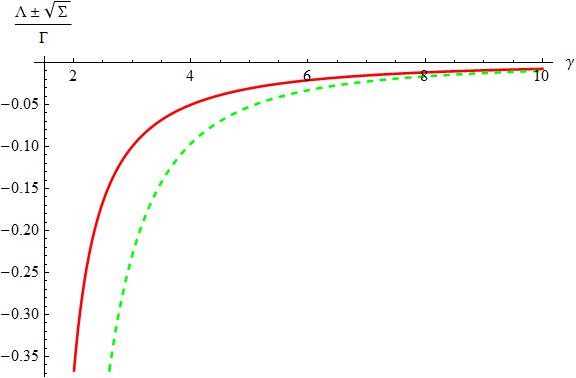}
\caption{The red and green (dashed) lines show the "+" and "-" branches, respectively.}
\label{fig_zero}
\end{figure}

As one can see in Figure~\ref{fig_zero}, there is no real solution of $\beta $ since the argument of the square root is always negative for $\gamma>3/2$. This is also the case when $\beta$ is extended to a complex parameter, because then the solution $\eqref{sol_zero}$ is  replaced by
\begin{align}
&|\beta|/M_{\rm{P}}= \sqrt{\frac{\Lambda \pm\sqrt{\Sigma}}{\Gamma}}
\end{align}
that cannot be imaginary.

However, for a non-vanishing $\epsilon$, the condition $\mathcal{V}(\beta,\gamma,\epsilon)=0$ can be satisfied. First, let us comment on the allowed regions of $\epsilon$. They are determined by requiring no ghost mode in the system. Since $\epsilon$ induces the kinetic mixing of $\varphi$ and $\Phi$ (see Eq.~$\eqref{Kahler}$), we have the following eigenvalues:
\begin{align}
\lambda_{\pm}=1+4\zeta |\Phi|^2\pm \sqrt{16\zeta^2 |\Phi|^4+\frac{\epsilon^2}{3}e^{2x}}  
\end{align}
in the diagonal basis of the kinetic terms. Therefore, to avoid a ghost mode during inflation, 
the $\epsilon$ must satisfy
\begin{align}
|\epsilon| <\sqrt{3}e^{-x}~,
\end{align}
where we have neglected the terms including $\Phi$, because they are suppressed by $\zeta$. At the point $x=5.5$, we have a constraint $|\epsilon|<\sqrt{3}e^{-x}\sim 7\times 10^{-3}$.

For example, given $\delta=10^{-5}$ and $\epsilon =-5.0\times 10^{-3}$,\footnote{In the case of 
$\epsilon> 0$, there is a singularity in the scalar potential.}  we obtain the two solutions numerically:
\begin{align}
&(i)\ \beta  \sim -76 M_{\rm{P}},\\
&(ii)\ \beta  \sim -266 M_{\rm{P}}.
\end{align}
In both cases, the masses of $\varphi, \Phi_R$ and $\Phi_I$ are estimated as 
\begin{align}
&(i)\ m^2_{\varphi}\sim 6.0\times 10^{-2}  \mu^2/M_{\rm{P}}^2,\ \ m^2_{R} \sim m^2_{I} \sim 2.1 \mu^2 |\zeta|,\\
&(ii)\ m^2_{\varphi}\sim 3.9 \mu^2/M_{\rm{P}}^2,\ \ m^2_{R} \sim m^2_{I} \sim 148 \mu^2 |\zeta|,
\end{align}
while they are all positive. 

To summarize this subsection, due to the existence of $\epsilon\neq 0$, we can find solutions of $\beta$ to get $\mathcal{V}=0$.~\footnote{We neglected the $\zeta$-suppressed terms in our analysis. When taking into account their contribution, our solutions for $\beta$ are going to be corrected by the $\mathcal{O}(1/\zeta)$ terms.} Then, the $D$-term contributions to the vacuum energy, neglected in Eqs.~$\eqref{EP}$ and $\eqref{vev_VF}$, become important. The vacuum energy  produced by $V_D$ is given by
\begin{align}
\nonumber  \langle V \rangle =&\langle V_D \rangle \\
 \sim &  \frac{M_{\rm{P}}^4}{a^4}\left(\sqrt{1+\frac{9}{4}a^4g^2\left(1-\langle e^{-x}\rangle-\frac{2}{3}\gamma -\frac{b}{9g}\right)^2}-1\right),
\end{align}
where $\langle x \rangle$ can be read off from  Eq.~$\eqref{vev_vacuum}$. 
To obtain a tiny positive cosmological constant, we have to tune the value of $b$. For example, by setting $b=9g\left(1-\langle e^{-x}\rangle-\frac{2}{3}\gamma\right) + \tilde{\epsilon}$, we find
\begin{align}
\langle V \rangle \sim   \frac{M_{\rm{P}}^4}{72}\tilde{\epsilon}^2~~.
\end{align}

To this end, we comment on the SUSY breaking scale in the vacuum. The vacuum expectation values of $F^{\Phi}$ and $D$ are given by
\begin{align}
&|\langle F^{\Phi}\rangle |\sim \mu e^{2\langle\mathcal{J}\rangle /M_{\rm{P}}^2}\frac{1}{\langle \Delta\rangle}\left(1-2 \epsilon \frac{\beta}{M_{\rm{P}}^2}\langle M_{\rm{P}}e^x+Q\rangle\right),\\
&|\langle D\rangle |\sim \frac{M_{\rm{P}}^2}{6}\tilde{\epsilon}~~.
\end{align}
As is clear from these equations, the F- and D- type breaking scales are decoupled, while
the $D$-type scale can be arbitrarily small.
In the examples (i) and (ii) above, the  $|\langle F^{\Phi}\rangle|$ is given by 
\begin{align}
&(i)\ \ |\langle F^{\Phi}\rangle |\sim 1.8\times 10^{4} \mu, \\
&(ii)\ \ |\langle F^{\Phi}\rangle |\sim 9.8\times 10^{-3} \mu. 
\end{align}

\section{Superspace actions and super-Higgs effect}

\newcommand{\overbar}[1]{\mkern1.5mu\overline{\mkern-1.5mu#1\mkern-1.5mu}\mkern 1.5mu}

It is instructive to reformulate our model in the curved superspace of the old-minimal supergravity, because it helps in construction of more general supergravity models, as
well as for verifying our calculations based on the superconformal tensor calculus.
  
\subsection{Superspace Lagrangian}

Let us consider the following superspace Lagrangian by using the standard notation and conventions of Ref.~\cite{Wess:1992cp},
\begin{gather}
    {\cal L}=\int d^2\Theta 2{\cal E}\left[\frac{3}{8}(\bar{{\cal D}}^2-8{\cal R})e^{-\frac{2}{3}{\cal J}}\left(1+\frac{4g{\cal I}}{9}\frac{{\cal W}^2\bar{{\cal W}}^2}{({\cal DW})^3}\right)+W(S_i)+L_{DBI}\right]+{\rm h.c.}\label{lagr1}
\end{gather}
Here ${\cal J}$ and ${\cal I}$ are arbitrary real functions of chiral superfields $S_i,\bar{S}_i$ and a real massive superfield $V$, whose gauge coupling is $g$; $W(S_i)$ is a holomorphic superpotential; ${\cal W}\equiv{\cal W}^\alpha{\cal W}_\alpha$ and ${\cal DW}\equiv {\cal D}^\alpha{\cal W}_\alpha$ where ${\cal W}_\alpha$ is the chiral superfield strength of $V$. The $L_{BI}$ is the DBI contribution,
\begin{equation}
    L_{DBI}=\frac{1}{4}f{\cal W}^2-\frac{\alpha}{16}(\bar{{\cal D}}^2-8{\cal R})\frac{|f+\bar{f}|^2{\cal W}^2{\bar{\cal W}}^2}{1+4\alpha(\omega+\bar{\omega})+\sqrt{1+8\alpha(\omega+\bar{\omega})+\frac{\alpha^2}{16}(\omega-\bar{\omega})^2}}~,
\end{equation}
where $f=f(S_i)$ is a holomorphic gauge kinetic function, $\alpha$ is the DBI parameter extended to a function of the chiral (matter) superfields, and $\omega$ is
\begin{equation}
    \omega=\frac{1}{16}(f+\bar{f})(\bar{{\cal D}}^2-8{\cal R}){\cal W}^2~.
\end{equation}

The Lagrangian \eqref{lagr1} is written after the superconformal gauge fixing in the so-called Jordan frame. After eliminating the auxiliary fields and the Weyl rescaling to the Einstein frame, the bosonic component Lagrangian reads
\begin{multline}
    e^{-1}{\cal L}=\frac{1}{2}R+\sqrt{1+8\alpha e^{-\frac{4}{3}{\cal J}}\frac{Z^2}{f_R}}\left(\frac{g\cal I}{6Z}-\frac{1}{4}\right)f_RF^2 +\frac{i}{4}f_IF\tilde{F}
    -\frac{1}{2}{\cal J}_{CC}\partial^aC\partial_aC  \\ -2{\cal J}_{i\bar{j}}\partial^aS^i\partial_a\bar{S}^j-\frac{1}{2}g^2{\cal J}_{CC}B^aB_a - {\cal J}_{Ci}\partial^aC\partial_aS^i-{\cal J}_{C\bar{i}}\partial^aC\partial_a\bar{S}^i \\
  +iB^a({\cal J}_{Ci}\partial_aS^i-{\cal J}_{C\bar{i}}\partial_a\bar{S}^i)-{\cal V}_D-{\cal V}_F+{\cal O}(F^4)~,
\end{multline}
with the F- and D-type scalar potentials
\begin{gather}
    {\cal V}_F=e^{2\cal J}\left[2\Delta^{i\bar{j}}\left(\chi_iW+\frac{1}{2}W_i\right)\left(\bar{\chi}_{\bar{j}}\bar{W}+\frac{1}{2}\bar{W}_{\bar{j}}\right)-\left(3-2\frac{{\cal J}_C^2}{{\cal J}_{CC}}\right)|W|^2\right]~,\label{VF}\\
    {\cal V}_D=\frac{e^{\frac{4}{3}{\cal J}}}{8\alpha}\left[\sqrt{1+8\alpha e^{-\frac{4}{3}{\cal J}}\frac{Z^2}{f_R}}-1\right]=\frac{Z^2}{2f_R}+{\cal O}(\alpha)~,\label{VD}
\end{gather}
where we have introduced the notation
\begin{gather}
    Z\equiv -g\left({\cal J}_C+\frac{\cal I}{6}\right)~,~~~f_R\equiv {\rm Re}f~,~~~f_I\equiv {\rm Im}f~,\nonumber\\
    \Delta^{i\bar{j}}\equiv\left({\cal J}_{i\bar{j}}-\frac{{\cal J}_{Ci}{\cal J}_{C\bar{j}}}{{\cal J}_{CC}}\right)^{-1}~,~~~\chi_i\equiv{\cal J}_i-\frac{{\cal J}_{Ci}{\cal J}_C}{{\cal J}_{CC}}~,~~~\bar{\chi}_{\bar{i}}\equiv {\cal J}_{\bar{i}}-\frac{{\cal J}_{C\bar{i}}{\cal J}_C}{{\cal J}_{CC}}~.
\end{gather}

The Lagrangian \eqref{lagr1} is more general than the one of Sec.~2 because it includes 
the gauge kinetic function $f(S_i)$. See Appendix A for more.

\subsection{The gauge-invariant reformulation}

The Lagrangian \eqref{lagr1} can be rewritten to the manifestly  $U(1)$ gauge-invariant
and manifestly supersymmetric form. Let us begin with another action
\begin{gather}
    {\cal L}=\int d^2\Theta 2{\cal E}\left[\frac{3}{8}(\bar{{\cal D}}^2-8{\cal R})e^{-\frac{1}{3}K+\Gamma}\left(1+\frac{4g{\cal I}}{9}\frac{{\cal W}^2\bar{{\cal W}}^2}{({\cal DW})^3}\right)+W(S_i)+L_{DBI}\right]+{\rm h.c.}\label{lagr2}
\end{gather}
where we have introduced the K\"ahler potential $K=K(T,\bar{T},S_i,\bar{S}_i)$ depending upon the Higgs/St\"uckelberg superfield $T$ and the neutral chiral superfields $S_i$. The $\Gamma=\Gamma(T,\bar{T},V)$ is the counterterm for the gauge transformations of $T$ (so that a sum $K+\Gamma$ is gauge invariant), while the vector gauge superfield $V$ is massless. Here the ${\cal I}$ is an arbitrary gauge-invariant real function ${\cal I}(T,\bar{T},S_i,\bar{S}_i,V)$. The $L_{DBI}$ is unchanged.

The linearly realized $U(1)$ gauge symmetry can be made explicit by requiring that the function $K+\Gamma$ only depends upon the product $Te^{2gV}\bar{T}$ that is invariant with respect to the gauge transformations
\begin{equation}
V\rightarrow V-\frac{1}{2}(\Lambda+\bar{\Lambda})~,~~~T\rightarrow Te^{g\Lambda}~,\label{gt1}
\end{equation}
where $\Lambda$ is an arbitrary chiral superfield. However, for our purposes, it is more convenient to let $T$ to transform under the $U(1)$ as follows:
\begin{equation}
V\rightarrow V-\frac{1}{2}(\Lambda+\bar{\Lambda})~,~~~T\rightarrow T-g\Lambda~,\label{gt2}
\end{equation}
so that the gauge-invariant combination is given by $T+\bar{T}-2gV$.

It is straightforward to calculate the bosonic terms of the Lagrangian \eqref{lagr2}.~\footnote{We use the same letters to denote the chiral superfields and their leading components.} We find
\begin{multline}
    e^{-1}{\cal L}=\frac{1}{2}R+\sqrt{1+8\alpha e^{-\frac{2}{3}K}\frac{Z^2}{f_R}}\left(\frac{g\cal I}{6\Tilde{Z}}-\frac{1}{4}\right)f_RF^2+\frac{i}{4}f_IF\tilde{F} \\
    -K_{i\bar{j}}\partial^aS^i\partial_a\bar{S}^j-K_{T\bar{T}}D^aT \overbar{D_aT}-
    K_{i\bar{T}}\partial^aS^i\overbar{D_aT}-K_{T\bar{i}}D^aT\partial_a\bar{S}^i \\
    -{\cal V}_D-{\cal V}_F+{\cal O}(F^4)~,
\end{multline}
where the $F$- and $D$-type scalar potentials are
\begin{gather}
    {\cal V}_F=e^K\left[K^{I\bar{J}}\left(K_IW+W_I\right)\left(K_{\bar{J}}\bar{W}+\bar{W}_{\bar{J}}\right)-3|W|^2\right]~,\label{VF2}\\
    {\cal V}_D=\frac{e^{\frac{2}{3}K}}{8\alpha}\left[\sqrt{1+8\alpha e^{-\frac{2}{3}K}\frac{\Tilde{Z}^2}{f_R}}-1\right]=\frac{\Tilde{Z}^2}{2f_R}+{\cal O}(\alpha)~.\label{VD2}
\end{gather}
Here $D_a$ is the $U(1)$ covariant derivative
\begin{equation}
    D_a\equiv\partial_a-gB_aX^T~,
\end{equation}
where $X^T$ is the Killing vector.~\footnote{The Killing vector of the linear gauge transformation \eqref{gt1} is $X^T=-iT$, whereas for the gauge transformation \eqref{gt2} we get $X^T=i$.} The indices $I,J$ in Eq.~\eqref{VF2} denote the scalar field $T$ and $S_i$ together, i.e. $I=\{i,T\}$. Since we do not consider the $U(1)$ gauge symmetry as the $R$-symmetry, the superpotential $W$ should be independent of $T$, i.e. $W_T=0$. The $\Tilde{Z}$ stands for
\begin{equation}
    \Tilde{Z}\equiv -g\left(\mathscr{D}+\frac{\cal I}{6}\right)~,
\end{equation}
where $\mathscr{D}$ is the Killing potential, which can be expressed as
\begin{equation}
    \mathscr{D}=iK_TX^T=-iK_{\bar{T}}X^{\bar{T}}~.
\end{equation}

A correspondence between the gauge-invariant Lagrangian \eqref{lagr2} and its "massive" counterpart \eqref{lagr1} can be established as follows. The "massive" description corresponds to the unitary gauge, $T=0$, in the convention \eqref{gt2}, of the Lagrangian \eqref{lagr2}. In this gauge, we choose $\Lambda=T/g$ in Eq.~\eqref{gt2}, which leads to
\begin{equation}
V\rightarrow \tilde{V}=V-\frac{1}{2g}(T+\bar{T}) \quad {\rm and} \quad T\rightarrow 0~.\label{gtt0}
\end{equation}

On the other hand, in the gauge-invariant formulation, we can use the Wess-Zumino gauge where the auxiliary components of $V$ are gauged away. In the leading order, this leads to
\begin{equation}
T+\bar{T}-2gC\rightarrow \tilde{T}+\bar{\tilde{T}}~,\label{gtwz}
\end{equation}
where $\tilde{T}=T-gC$ is the St\"uckelberg scalar that absorbs the auxiliary real scalar $C$. Since the left-hand-side of Eq.~\eqref{gtt0} in the leading order reads
 $T+\bar{T}-2gC\rightarrow -2g\tilde{C}$, comparing it with Eq.~\eqref{gtwz} yields the relation
\begin{equation}
2g\tilde{C}=-\tilde{T}-\bar{\tilde{T}}~.\label{ctt}
\end{equation}
Thus we can easily switch between the Lagrangian \eqref{lagr1} in terms of ${\cal J}(C,...)$ and the Lagrangian \eqref{lagr2} in terms of $K(T,\bar{T},...)$.

As an example, let us consider the Starobinsky case with $\cal J$ in Eq.~ \eqref{lagr1} given by \cite{Ferrara:2013rsa}
\begin{equation}
    2{\cal J}=-3\log(-C)-3C~.\label{Jstar}
\end{equation}
This can be translated into the gauge-invariant notation by sending $C\rightarrow 2gC$ in Eq.~\eqref{Jstar} and using Eq.~\eqref{ctt}. The resulting theory is now  described by the Lagrangian \eqref{lagr2} with~\footnote{The gauge couplings $g$ in the two formulations are related to each other by the factor of two, i.e. one should take $g\rightarrow g/2$ in 
Eq.~\eqref{lagr2}, or $g\rightarrow 2g$ in Eq.~\eqref{lagr1}.}
\begin{equation}
    K=-3\log(T+\bar{T})+3(T+\bar{T})~,
\end{equation}
whose first term corresponds to the no-scale supergravity.

\section{Conclusion}

Our purpose and motivation are to {\it unify} (i) Starobinsky inflation, (ii) high-scale spontaneous SUSY breaking after inflation, and (iii) a positive cosmological constant, in supergravity. To achieve the goal (i), we employ the {\it minimal} number of the physical degrees of freedom contained in the single massive vector multiplet: the inflaton, the goldstino, and the massive vector field. The possible observational signatures of the massive vector field may arise as non-gaussianities of the CMB spectrum 
\cite{Arkani-Hamed:2015bza}. To achieve the goal (ii), needed for reheating and viable particle phenomenology, we employ the {\it minimal} hidden sector described by the single chiral (Polonyi) multiplet. The goal (iii) is needed to describe the dark energy (the accelerating Universe) via a de Sitter vacuum. 

Despite the considerable progress, mentioned in our references and in the main text, this line of research faced several problems, such as (a) instability of the inflationary trajectory due to the mixing of the inflaton and Polonyi scalar, (b) the limited use of the
standard FI term in the supergravity-based cosmology, and  (c) the problem of decoupling the F-type and D-type contributions to the scalar potential, needed for the hierarchy between the observed cosmological constant and a much higher SUSY breaking scale.

Overcoming these problems is apparently possible only with additional theoretical resources
provided by the DBI structure and the new (generalized) FI terms that do not require the R-symmetry gauging and, hence, allow for more general couplings.

We demonstrated in this paper that the goals (i), (ii) and (iii) can be {\it simultaneously} achieved, though in the rather complicated way, via the extension of the original Polonyi model and the tuning of the parameters.

Our model after stabilization of the Polonyi scalar by construction leads to {\it the same} predictions for the CMB observables as the Starobinsky model due to the choice \eqref{Starobinsky} of the $J$-function and the chosen value of the coupling constant $g$ in Subsect.~4.1. 
As regards the tiny value of the cosmological constant  $\Lambda^{1/4} \approx 0.0024$ eV, it arises
entirely due to the D-term by demanding (fine-tuning) the vanishing F-type term, i.e. the vanishing difference between the two relatively large contributions with the opposite signs in Eq.~\eqref{vev_VF}.

As regards the so-called Polonyi problem (i.e. possible overproduction of Polonyi particles in the 
early Universe after inflation), it may be avoided in the supergravity model under consideration
\cite{Addazi:2017ulg}. 

We used the more general Polonyi couplings demanded by consistency of our approach. 
It is still desirable to fix interactions of the hidden (Polonyi) sector by some organizing (symmetry) principles beyond local supersymmetry. It could be S-duality or extra (hidden and non-linearly realized) supersymmetry of the relevant actions in the context of {\it partial} SUSY breaking, e.g., by exploiting possible connections to $D3$-branes and their anti-branes in string theory  \cite{Cribiori:2018dlc}, though this is  beyond the scope of this investigation.

\newpage

\section*{Acknowledgements}

HA was supported in part by the JSPS (kakenhi) Grant under No.JP16K05330 (HA). YA was supported by the CUniverse research promotion project of Chulalongkorn University in Thailand under the grant reference CUAASC, and the Ministry of Education and Science of the Republic of Kazakhstan under the grant reference AP05133630. SA was supported in part by Waseda University Grant for Special Research Projects under No.~2018S-141. 
SVK was supported in part by the Grant-in-Aid of the Japanese Society for Promotion of Science (JSPS) under No.~26400252 (SK),  the Competitiveness Enhancement Program of Tomsk Polytechnic University in Russia,  and the World Premier International Research Center Initiative (WPI Initiative), MEXT, Japan. 


\begin{appendix} 

\section{Extended DBI sector} \label{detail_DBI}

Let us study some matter coupled generalizations of the DBI sector, $S_{DBI}$, in the
superconformal setting. These extensions can be found, e.g., in Refs.~\cite{Abe:2015nxa,Aoki:2016tod}, where the constraint related to the partial breaking of 
$N=2$ supersymmetry was used. As a specific extension of the DBI sector,
let us take

\begin{align}
S_{DBI}=&-\frac{1}{2}[fW^2]_F+\biggl[  \frac{\alpha \Psi W^2\bar{W}^2}{|S_0|^4\left( 1-2\alpha \mathcal{A}+\sqrt{1-4\alpha \mathcal{A}+4\alpha ^2\mathcal{B}^2} \right)}\biggr] _D, \label{S_DBI2}
\end{align} 
where 
\begin{align}
\mathcal{A}=\bar{\Sigma }\left( \frac{W^2}{|S_0|^4}\Omega \right) +{\rm{h.c.}}, \ \ \mathcal{B}=\bar{\Sigma }\left( \frac{W^2}{|S_0|^4}\Omega \right) -{\rm{h.c.}}~,
\end{align} 
$f$ is a holomorphic gauge kinetic function of $S^i$. In general, $\alpha$ (real), $\Psi$ (real) and $\Omega$ (complex) depend on $V,S^i$ and $\bar{S}^{\bar{i}}$. All of them are the weightless multiplets. The bosonic expansion of Eq.~$\eqref{S_DBI2}$ reads
\begin{align}
\nonumber \mathcal{L}_{DBI}=&-\frac{1}{4}\left(f_R-\frac{\Psi \Omega_R}{|\Omega|^2}\right)FF+\frac{i}{4}\left(f_I-\frac{\Psi\Omega_I}{|\omega|^2}\right)F\tilde{F}\\
\nonumber &+\frac{1}{2}\left(f_R-\frac{\Psi\Omega_R}{|\Omega|^2}\right)D^2\\
\nonumber &+\frac{|S_0|^4\Psi}{8\alpha|\Omega|^2}\biggl[ 1- \biggl\{1+\frac{4\alpha}{|S_0|^4}\left( -2\Omega_RD^2+\Omega_RFF-i\Omega_IF\tilde{F}\right)\\
&+\frac{4\alpha^2}{|S_0|^8}\left( -2i\Omega_ID^2+i\Omega_IFF-\Omega_RF\tilde{F}\right)^2\biggr\}^{1/2}\biggr] ,\label{expansion}
\end{align} 
where we have used the same letters to denote the first components of $f, \alpha, \Psi$ and $\Omega$ as for the corresponding multiplets or superfields. In addition, the "$R$" and "$I$" subscripts denote the real and imaginary parts. 

First, we demand the above Lagrangian to have the DBI form (for $F_{ab})$ when $D=0$,
by using the identity
\begin{align}
-{\rm{det}}\left(   \eta_{ab}+a   F_{ab} \right) =1+\frac{a ^2}{2}FF+\frac{a ^4}{16}(F\tilde{F})^2~. \label{DBI_formula}
\end{align}
This gives rise to the following consequences:
\begin{enumerate}
  \item The first term in Eq.~$\eqref{expansion}$ should vanish, because Maxwell term comes from the expansion of the DBI term in the weak field limit. On the other hand, the $F\tilde{F}$ does not come from the DBI expansion, so that the second term in 
  Eq.~$\eqref{expansion}$ should not vanish.
  \item Inside the square root (focusing on the third and fourth lines of Eq.~$\eqref{expansion}$), we need $FF$, but not $(FF)^2$. In the same way, we need $(F\tilde{F})^2$, but not $F\tilde{F}$. Otherwise, the expansion formula $\eqref{DBI_formula}$ can never be applied.
\end{enumerate}
After taking the both consequences into account, we obtain
\begin{align}
\Omega_R=\frac{\Psi}{f_R} \quad {\rm and} \quad  \Omega_I=0~.
\end{align}
For example, the case 
\begin{align}
\Psi =f_R^2
\end{align}
leads to
\begin{align}
\mathcal{L}_{DBI}=&\frac{i}{4}f_IF\tilde{F}+\frac{|S_0|^4}{8\alpha}\biggl[ 1- \sqrt{-{\rm{det}}\left(g_{\mu \nu}+\sqrt{\frac{8\alpha f_R}{|S_0|^4}}F_{\mu\nu}\right)}\biggr]~. \label{DBIform1}
\end{align}
Hence, when $D\neq 0$, as in our case, the DBI structure becomes generically more complicated.


\section{Derivation of $S_{V,M}$} \label{detail_VM}

The bosonic expansion of Eq.~$\eqref{S_VM}$ reads
\begin{align}
\nonumber  \mathcal{L}_{V,M}=&-\frac{1}{3}|S_0|^2\mathcal{H}R(\omega)+2\mathcal{H}\left( |F_0|^2-|D_aS_0|^2\right) +|S_0|^2\mathcal{H}_CD\\
\nonumber &+\frac{1}{2}|S_0|^2\mathcal{H}_{CC}\left( |N|^2-B_a^2-(D_aC)^2\right) \\
\nonumber &+2|S_0|^2\mathcal{H}_{i\bar{j}}\left( F^i\bar{F}^{\bar{j}}-\partial_aS^i\partial^a\bar{S}^{\bar{j}}\right) \\
\nonumber &+\biggl\{-\mathcal{H}_CNS_0\bar{F}_0+\mathcal{H}_Ci\left( B_a+iD_aC\right) S_0D^a\bar{S}_0\\
\nonumber &-|S_0|^2\mathcal{H}_{Ci}\left(F^i\bar{N}+i\partial_aS^i(B^a-iD^aC)\right)\\
&+2S_0\mathcal{H}_i\bar{F}_0F^i-2S_0\mathcal{H}_iD_a\bar{S}_0\partial^aS^i+3S_0^2F_0\mathcal{W}+S_0^3F^i\mathcal{W}_i+{\rm{h.c.}} \biggr\} , \label{L_VM1}
\end{align}
where $N\equiv H+iK$. The subscripts on $\mathcal{H}$ and $\mathcal{W}$ denote the derivatives with respect to $C, S^i$ and $\bar{S}^{\bar{i}}$, e.g., $\mathcal{H}_C=\partial\mathcal{H}/\partial C$ and $\mathcal{H}_i=\partial\mathcal{H}/\partial S^i$. The $D_a$ is the superconformal covariant derivative~\cite{Freedman:2012zz}, whose  bosonic part is given by
\begin{align}
D_aS_0=\partial_a S_0-iA_aS_0-b_aS_0, \ \ D_aC=\partial_a C-2b_aC.
\end{align}

We take the following superconformal gauge fixing conditions:
\begin{align}
&{\rm{D-gauge}} :\ -\frac{1}{3}|S_0|^2\mathcal{H}=\frac{1}{2} ~,\label{D_gauge}\\
&{\rm{A-gauge}} :\ S_0=\bar{S}_0~,\label{A_gauge}\\
&{\rm{K-gauge}} :\ b_{\mu}=0~,\label{K_gauge}
\end{align}
which ensure that the Ricci scalar is canonically normalized in the action.~\footnote{The gauge fixing condition for special SUSY is irrelevant for the bosonic part.} Then the $R(\omega)$ becomes the usual Ricci scalar that we denote by $R$. Under these gauge conditions, we can rewrite Eq.~$\eqref{L_VM1}$ as 
\begin{align}
\nonumber \mathcal{L}_{V,M}&=\frac{1}{2}R+2\mathcal{H}|F_0|^2-\frac{3}{4\mathcal{H}^2}I_a^2 +\frac{3}{\mathcal{H}}A_aJ^a+3A_a^2\\
\nonumber &-\frac{3}{2\mathcal{H}}\biggl\{\mathcal{H}_CD+\frac{1}{2}\mathcal{H}_{CC}\left(|N|^2-B_a^2-(\partial_aC)^2\right)+2\mathcal{H}_{i\bar{j}}\left( F^i\bar{F}^{\bar{j}}-\partial_aS^i\partial^a\bar{S}^{\bar{j}}\right)\biggr\} \\
\nonumber &+\biggl\{\frac{3\mathcal{H}_{Ci}}{2\mathcal{H}}\left(F^i\bar{N}+i\partial_aS^i(B^a-iD^aC)\right)+\sqrt{\frac{-3}{2\mathcal{H}}}\left(-\mathcal{H}_CN\bar{F}_0+2\mathcal{H}_i\bar{F}_0F^i\right)\\
&-\frac{9}{2\mathcal{H}}F_0\mathcal{W}+\left(\frac{-3}{2\mathcal{H}}\right)^{3/2}F^i\mathcal{W}_i+{\rm{h.c.}} \biggr\} ~~, \label{L_VM2}
\end{align}
where we have used the notation
\begin{align}
&I_a=\mathcal{H}_C\partial_aC+\mathcal{H}_i\partial_aS^i+\mathcal{H}_{\bar{j}}\partial_a\bar{S}^{\bar{j}},\\
&J_a=\mathcal{H}_CB_a+i\mathcal{H}_i\partial_aS^i-i\mathcal{H}_{\bar{j}}\partial_a\bar{S}^{\bar{j}}.
\end{align}

Next, the auxiliary fields $A_a,F_0,F^i,N$ and $D$ should be eliminated. As was already mentioned in the main text, only $D$ non-trivially appears in $\mathcal{L}_{DBI}$ and $\mathcal{L}_{FI}$, and therefore, we can easily integrate over the other auxiliary fields. Their algebraic solutions are 
\begin{align}
&A_a=-\frac{1}{2\mathcal{H}}J_a,\\
&F^i=-e^{\mathcal{J}}\Delta^{i\bar{j}}\biggl\{\frac{1}{2}\bar{\mathcal{W}}_{\bar{j}}+\left(\mathcal{J}_{\bar{j}}-\frac{\mathcal{J}_{C\bar{j}}\mathcal{J}_C}{\mathcal{J}_{CC}}\right)\bar{\mathcal{W}}\biggr\},\\
&N=\frac{2}{\mathcal{J}_{CC}}\left(\mathcal{J}_{Ci}F^i+e^{\mathcal{J}}\mathcal{J}_C\bar{\mathcal{W}}\right),\\
&F_0=\frac{2}{3}e^{\mathcal{J}/3}\left(\mathcal{J}_i-\frac{\mathcal{J}_{Ci}\mathcal{J}_C}{\mathcal{J}_{CC}}\right)F^i-e^{4\mathcal{J}/3}\left(\frac{2\mathcal{J}_C^2}{3\mathcal{J}_{CC}}-1\right)\bar{\mathcal{W}},
\end{align}
where we have defined 
\begin{align}
&\mathcal{J}=-\frac{3}{2}\log \left( -\frac{2}{3}\mathcal{H}\right),\\
&\Delta^{i\bar{j}}=\left(\Delta_{i\bar{j}}\right)^{-1}, \ \ \Delta_{i\bar{j}}=\mathcal{J}_{i\bar{j}}-\frac{\mathcal{J}_{Ci}\mathcal{J}_{C\bar{j}}}{\mathcal{J}_{CC}}~.
\end{align}
After substituting these solutions into Eq.~$\eqref{L_VM2}$, we finally obtain
\begin{align}
\nonumber \mathcal{L}_{V,M}&=\frac{1}{2}R-\frac{1}{2}\mathcal{J}_{CC}(\partial_aC)^2 -\frac{1}{2}\mathcal{J}_{CC}B_a^2+\mathcal{J}_CD\\
\nonumber &-2\mathcal{J}_{i\bar{j}}\partial_aS^i\partial^a\bar{S}^{\bar{j}}+\left(-\mathcal{J}_{Ci}\partial_aC\partial^aS^i-i\mathcal{J}_{Ci}B_a\partial^aS^i+{\rm{h.c.}}\right)\\
\nonumber &-e^{2\mathcal{J}}\biggl[2\Delta^{i\bar{j}}\biggl\{\frac{1}{2}\mathcal{W}_{i}+\left(\mathcal{J}_{i}-\frac{\mathcal{J}_{Ci}\mathcal{J}_C}{\mathcal{J}_{CC}}\right) \mathcal{W}\biggr\}\biggl\{\frac{1}{2}\bar{\mathcal{W}}_{\bar{j}}+\left(\mathcal{J}_{\bar{j}}-\frac{\mathcal{J}_{C\bar{j}}\mathcal{J}_C}{\mathcal{J}_{CC}}\right) \bar{\mathcal{W}}\biggr\}\\
&+2\frac{\mathcal{J}_C^2}{\mathcal{J}_{CC}}|\mathcal{W}|^2-3|\mathcal{W}|^2\biggr] ~. \label{L_VM3}
\end{align}

\section{Derivation of the coefficients in Eq.~$\eqref{EP}$} \label{explicit_mass}

Here we compute the explicit coefficients of Eq.~$\eqref{EP}$. We skip writing down  
$\delta_R$ and $m^2_{R\varphi}$, whose explicit expressions are not used in the main text. 

First, the $V_0$ is given by
\begin{align}
\nonumber V_0=&\frac{e^{3-2 \gamma } \mu ^2}{3M_{\rm{P}}^2(3-4\epsilon)} \biggl[9M_{\rm{P}}^2-12 M_{\rm{P}} \beta (3+2\gamma)\epsilon +\beta^2(3+2\gamma) (-9+12\epsilon+12\epsilon^2\\
 &+6\gamma+8\gamma \epsilon (\epsilon-1))\biggr]. \label{CC}
\end{align}

The linear term $\delta_{\varphi}$ reads
\begin{align}
\nonumber \delta_{\varphi}=&\frac{2\mu^2}{3 M_{\rm{P}}^3 (3-4 \epsilon )^2} \sqrt{\frac{2}{3}} e^{3-2 \gamma }\biggl[ 9 M_{\rm{P}}^2 (4 \gamma \epsilon +4\epsilon -3 \gamma )+\beta ^2 \biggl\{  4 \gamma ^3 (4 \epsilon -3) (4 \epsilon -4 \epsilon^2 +3)\\
\nonumber &+4 \gamma ^2 \left(4 \epsilon  \left(8 \epsilon ^2+\epsilon -6\right)+9\right)+3 \gamma  (4 \epsilon  (12 \epsilon^2 +7\epsilon-6)+9)+144 \epsilon ^3\biggr\}\\
&-12 \beta  M_{\rm{P}} \epsilon  \left(2\gamma^2(-3+4\epsilon)+3\gamma(-1+4\epsilon)+12\epsilon \right)\biggr]~.
\end{align}

The $m_{\varphi}^2$ is given by the following complicated expression:
\begin{align}
\nonumber m^2_{\varphi}=&-\frac{2 \mu^2e^{3-2 \gamma }}{9 M_{\rm{P}}^4 (4 \epsilon -3)^3}\Biggl[12 \beta  M_{\rm{P}} \epsilon  \biggl\{8\gamma^3(-16\epsilon^2+24\epsilon-9)+8\gamma^2 (-16\epsilon^2+9)\\
\nonumber &+24\gamma (-16\epsilon^2+8\epsilon-3)-3(80\epsilon^2+24\epsilon+9)\biggr\}\\
\nonumber &+9 M_{\rm{P}}^2 \left(4 \gamma ^2 (3-4 \epsilon )^2+6 \gamma  (4 \epsilon -3) (4 \epsilon +1)+8 \epsilon  (14 \epsilon -3)+27\right)\\
\nonumber &+\beta ^2 \biggl\{16 \gamma ^4 (3+4 \epsilon -4 \epsilon^2) (3-4 \epsilon )^2+8 \gamma ^3 (4 \epsilon -3) (4\epsilon^2 (29+4\epsilon)-120\epsilon +45)\\
\nonumber &+8 \gamma ^2 (4\epsilon^2(171-80\epsilon+80\epsilon^2)-540\epsilon+135)\\
\nonumber &+6 \gamma  (4\epsilon^2(-45+24\epsilon+176\epsilon^2)+324\epsilon-81)\\
&+9 (4\epsilon^2(9+4\epsilon)(-5+12\epsilon)+108\epsilon-27)\biggr\}\Biggr]. \label{mass_inflaton}
\end{align}

Finally, as regards the $m^2_{R}$ and the $m^2_{I}$, we only show their dominant terms with respect to $\zeta$ as follows:
\begin{align}
m^2_{R}=&m^2_{I}=-\frac{8\zeta \mu^2}{M_{\rm{P}}^2 (3-4 \epsilon )^2}e^{3-2 \gamma }  (3 M_{\rm{P}}-2 \beta  (2 \gamma +3) \epsilon )^2~.
\end{align}


\section{Minkowski vacuum after inflation}\label{Minkvac}

Let us discuss in more detail the possibility of Minkowski vacuum for the model described in Sec. \ref{ISB} with $\epsilon=\zeta=0$.
Consider the superspace action \eqref{lagr1} with $f=1$ and a constant FI parameter $\cal I$ (we take $M_P=1$ below):
\begin{align}
    2{\cal J}&=A\log(-C)-BC+\Phi\bar{\Phi}+d~,\label{J}\\
    W&=\mu(\Phi+\beta)~,\label{W}
\end{align}
where $C$ is the real scalar component $V$, $\Phi$ is the Polonyi complex scalar, and  $A,B,d,\mu,\beta$ are real constant parameters. This choice is related to the $\cal J$-function of Sec.~\ref{ISB} by setting $A=-3$ and $B=3-2\gamma$.

The choice \eqref{J} and \eqref{W} leads to the scalar potentials
\begin{gather}
    {\cal V}_F=\mu^2(-C)^Ae^{\Phi\bar{\Phi}-BC+d}\left(|\Phi\bar{\Phi}+\beta\Phi+1|^2-X|\Phi+\beta|^2\right)~,\label{vf}\\
    {\cal V}_D=\frac{g^2}{8}\left(B-\frac{A}{C}-\frac{\cal I}{3}\right)^2+{\cal O}(\alpha)~,\label{vd}\\
    X\equiv 3+A-2BC+\frac{B^2C^2}{A}~.
\end{gather}

Requiring the canonical kinetic term for the inflaton yields $C=\lambda e^{\sigma\varphi}$, with $\sigma=\sqrt{-2/A}$. Using the notation of Ref.~\cite{Abe:2018plc}, and can set 
$\lambda=-1$ without loss of generality because $\lambda$ can be absorbed by $B$ and $d$ in Eq.~\eqref{J}. During inflation, unless $B\leq 0$, the double exponential factor creates inflationary instability \cite{Aldabergenov:2017hvp}. In addition, the absence of ghosts requires $A<0$. Taking all this into account, it follows from Eq.~\eqref{VD} that, in order to obtain the Starobinsky inflationary potential, we must require 
$B-{\cal I}/3>0$, or ${\cal I}<-3|B|$.

The Minkowski vacuum conditions for the potential of Eqs.~\eqref{vf} and \eqref{vd} read
\begin{align}
\nonumber    {\cal V}&=\mu^2(-C_0)^Ae^{\Phi_0^2-BC_0+d}\left[(\Phi_0^2+\beta\Phi_0+1)^2-X_0(\Phi_0+\beta)^2\right] \\
    &    +\frac{g^2}{8}\left(B-\frac{A}{C_0}-\frac{\cal I}{3}\right)^2=0~,\label{vac1}\\
\nonumber    {\cal V}_{\Phi}&=\Phi_0{\cal V}_F+\mu^2(-C_0)^Ae^{\Phi_0^2-BC_0+d}\times \\
& \times     \left[(2\Phi_0+\beta)(\Phi_0^2+\beta\Phi_0+1)-X_0(\Phi_0+\beta)\right]=0~,\label{vac2}\\
 \nonumber  {\cal V}_\varphi&=\frac{\sigma A}{C_0}\left(B-\frac{A}{C_0}-\frac{\cal I}{3}\right)+\sigma(A-BC_0){\cal V}_F \\
 &   +2\sigma\mu^2(-C_0)^Ae^{\Phi_0^2-BC_0+d}(\Phi_0+\beta)^2BC_0\left(1-\frac{BC_0}{A}\right)=0~,\label{vac3}
\end{align}
where the subscript $0$ denotes the vacuum expectation values.

Let us consider the case when ${\cal V}_D$ and ${\cal V}_F$ separately vanish at the minimum.~\footnote{In this case, the DBI corrections can be dropped because they are all proportional to powers of $(B-\frac{A}{C}-\frac{{\cal I}}{3})$.} Then, as regards $C_0$, we have
\begin{equation}
    C_0=-e^{\sigma\varphi_0}=\frac{A}{B-\frac{{\cal I}}{3}}~~,
\end{equation}
while Eqs.~\eqref{vac1}-\eqref{vac3} are reduced to
\begin{gather}
    (\Phi_0^2+\beta\Phi_0+1)^2=X_0(\Phi_0+\beta)^2~,\label{vacr1}\\
    (2\Phi_0+\beta)(\Phi_0^2+\beta\Phi_0+1)=X_0(\Phi_0+\beta)~,\label{vacr2}\\
    (\Phi_0+\beta)^2BC_0\left(1-\frac{BC_0}{A}\right)=0~,\label{vacr3}
\end{gather}
respectively. 

As is clear from Eq.~\eqref{vacr3}, if we assume a non-vanishing $B$, then we get $B,C_0,A<0$, and the only way to satisfy Eq.~\eqref{vacr3} is to require $\Phi_0+\beta=0$ that is inconsistent with Eq.~\eqref{vacr1}. Hence, there is no Minkowski vacuum (at least for the case ${\cal V}_D={\cal V}_F=0$) unless $B=0$ (i.e. $\gamma=3/2$).

Given  $B=0$, we have
\begin{equation}
    X_0=X=3+A \quad {\rm and} \quad C_0=-\frac{3A}{\cal I}~,
\end{equation}
and the solutions to Eqs.~\eqref{vacr1}\eqref{vacr2} are given by
\begin{gather}
    {\rm (a)}~~\Phi_0=\sqrt{3+A}\pm 1~,~~~\beta=-\sqrt{3+A}\mp 2~,\label{sola}\\
    {\rm (b)}~~\Phi_0=-\sqrt{3+A}\pm 1~,~~~\beta=\sqrt{3+A}\mp 2~.\label{solb}
\end{gather}
Since $A$ must be negative, there is the additional restriction $-3\leq A<0$. Our choice of Sec. \ref{ISB}, $A=-3$, is the lowest allowed value in this case.

To summarize, we began with 7 parameters $A,B,d,g,\mu,\beta,{\cal I}$. Requiring the Starobinsky inflationary potential and Minkowski vacuum with spontaneously broken SUSY after inflation leaves only two parameters $g$ and $\mu$. The $B$ should vanish, and  the $d$ can also vanish because it does not have a significant impact. The orders of magnitude of $A$ and $\beta$ are also fixed, whereas the $\cal I$ can be absorbed into a redefinition of $\varphi$ and $g$. The gauge coupling $g$ controls the mass of the vector multiplet, including the mass of the inflaton $\varphi$, while the parameter $\mu$ is proportional to the gravitino mass and remains arbitrary in our models.

\end{appendix}


\end{document}